\documentclass[preprint,aps,pra,showpacs,floatfix]{revtex4}
\usepackage{graphicx}
\usepackage{times}
\usepackage{nicefrac}
\usepackage{amsmath}
\usepackage{amsfonts}
\usepackage{amssymb}
\usepackage{amsthm}
\usepackage{epsf}
\usepackage{bm}
\usepackage{bbm}

\usepackage{dcolumn}
\newcolumntype{.}{D{x}{}{-1}}

\newcommand{\be}{\begin{eqnarray}}
\newcommand{\ee}{\end{eqnarray}}
\newcommand{\la}{\langle}
\newcommand{\ra}{\rangle}

\newcommand{\veps}{\varepsilon}

\newcommand{\pr}{\prime}
\newcommand{\balpha}{\bm{\alpha}}
\newcommand{\bnabla}{\bm{\nabla}}
\newcommand{\bsigma}{\bm{\sigma}}

\newcommand{\bgamma}{\bm{\gamma}}
\newcommand{\bGamma}{\bm{\Gamma}}

\newcommand{\bmu}{\bm{\mu}}

\newcommand{\rmd}{{\rm d}}
\newcommand{\bfk}{{\bf k}}

\newcommand{\bfp}{{\bf p}}
\newcommand{\bfpr}{{\bf p^\prime}}

\newcommand{\np}{\not{p}}
\newcommand{\npr}{\not{p^\prime}}
\newcommand{\nk}{\not{k}}
\newcommand{\rp}{p_r}
\newcommand{\hp}{{\bf \hat p}}
\newcommand{\chid}{\chi^\dagger}
\newcommand{\bfr}{{\bf r}}

\newcommand{\dddd}{\rmd^4}
\newcommand{\dbp}{\rm{d \bfp}}
\newcommand{\dbpr}{\rm{d \bfpr}}

\newcommand{\dirr}{\delta^{\rm irr}}
\newcommand{\dvrz}{\delta^{{\rm vr}(0)}}
\newcommand{\dvro}{\delta^{{\rm vr}(1)}}
\newcommand{\dvrm}{\delta^{{\rm vr}(2+)}}
\newcommand{\dvrom}{\delta^{{\rm vr}(1+)}}
\begin{document}

\title{Radiative corrections to the magnetic-dipole 
transition amplitude in B-like ions}
\author{A. V. Volotka,$^{1,2}$ D. A. Glazov,$^{2}$ G. Plunien,$^{1}$
V. M. Shabaev,$^{2}$ and I. I. Tupitsyn$^{2}$}

\affiliation{
$^1$ Institut f\"ur Theoretische Physik, Technische Universit\"at Dresden,
Mommsenstra{\ss}e 13, D-01062 Dresden, Germany \\
$^2$ Department of Physics, St. Petersburg State University,
Oulianovskaya 1, Petrodvorets, 198504 St. Petersburg, Russia \\
}

\begin{abstract}
The one-electron quantum-electrodynamic corrections to the
magnetic-dipole transition amplitude 
between the fine-structure
levels $(1s^2 2s^2 2p) \, ^2P_{3/2} \, - \, ^2P_{1/2}$ in boronlike ions
are calculated to all orders in $\alpha Z$. The results obtained
serve for improving the theoretical accuracy of the lifetime of the
$(1s^2 2s^2 2p) \, ^2P_{3/2}$ level in boronlike argon.
\end{abstract}


\maketitle

%
\section{Introduction}
The precision in measurements of decay rates of forbidden transitions
has considerably increased during the last years
\cite{moehs98,traebert99,traebert00,traebert01,traebert02,lapierre05,lapierre06}.
The accuracy achieved for the magnetic-dipole (M1) transition
$(1s^2 2s^2 2p) \, ^2P_{3/2} \, - \, ^2P_{1/2}$ in B-like Ar became better
than one part per thousand \cite{lapierre05,lapierre06}. This experimental precision
demands a corresponding increase in the accuracy of the theoretical
predictions. 
In a recent work \cite{tupitsyn05} we have calculated the
M1-transition probabilities between the fine-structure levels in B-
and Be-like ions within the region of nuclear charge numbers $Z=16-22$.
In particular, it was found that the theoretical result
for the case of Ar$^{13+}$ deviates from the experimental one by about
$3\sigma$.
 
In Ref.~\cite{tupitsyn05}, the relativistic, interelectronic-interaction,
and quantum-electrodynamic (QED) corrections to the M1-transition amplitude
were computed, while experimental values were taken for the transition
energies. The configuration-interaction method in the Dirac-Fock-Sturm
basis (CIDFS) was employed in order to evaluate the interelectronic-interaction contribution.
Corrections due to single excitations to the negative-continuum energy states were taken
into account in the many-electron wave functions. 
As it is known from \cite{indelicato96,derevianko98}, such corrections
may be significant in calculations involving operators, which mix large and small
components of the wave functions, such as the M1-transition operator.
The frequency-dependent term (consult the detailed description presented
in Ref.~\cite{indelicato04}) was calculated within perturbation theory to first
order in $1/Z$. The QED correction was obtained within leading order by including
the electron anomalous magnetic moment (EAMM) in the M1-transition operator.
Uncalculated higher-order QED terms together with the experimental
errors of the transition energy determine the total uncertainty of the
theoretical predictions presented in Ref. \cite{tupitsyn05}. 
In this work, which is aiming for improvements of
the evaluation of the radiative effects to the M1-transition
probability in B-like ions, we present the exact calculation of the one-electron
QED corrections going beyond the EAMM approximation.

Accordingly, the bound-electron propagator is treated exactly. This approach
was already employed for the evaluation of the
radiative corrections to the decays $2p_{1/2},2s,2p_{3/2}-1s$ in hydrogenic
ions \cite{sapirstein04} and parity nonconserving transitions in neutral
Cs and Fr \cite{shabaev05a,shabaev05b}. Besides, in Ref.  \cite{shabaev98}
the QED corrections to  the transition
probability between the hyperfine-structure components were expressed
in terms of the corresponding corrections to 
 the bound-electron
g factor. The latter ones were calculated to all orders in $\alpha Z$
for the $1s$ and $2s$ states 
in Refs.~\cite{blundell97,persson97,beier00,yerokhin02,yerokhin04}.

Relativistic units ($\hbar=c=m=1$) and the Heaviside charge unit
[$\alpha=e^2/(4\pi), \, e<0$] are used throughout the paper.
%
\section{Basic formulas}
\label{sec:bf}
%
The magnetic-dipole transition probability between the one-electron
states $a$ and $b$ can be written in the form
\be
  W_{\rm D} = \frac{2\pi}{2 j_a+1} \,
  \sum_{m_a} \, \sum_{m_b} \, \sum_M \, |A_{1M}|^2 \,,
\ee
where the summation over the photon polarization and the integration
over the photon energy and angles were carried out. The initial state $a$
is characterized by the angular momentum $j_a$, its projection $m_a$, and
the energy $\veps_a$, while the final state $b$ has the corresponding
quantum numbers $j_b$, $m_b$, and the energy $\veps_b$.
The transition amplitude $A_{1M}$ is defined by 
\be
  A_{1M} = -\sqrt{\frac{3 \, \omega}{\pi}} \,
  \la b| T^1_M |a\ra \,,
\label{ampl}
\ee
where $\omega$ is the transition energy and $T^1_M$ denote
the spherical components of the M1-transition operator
\be
  {\bf T}^1 = \frac{e}{\sqrt{2}} \, j_1(\omega r) \,
  \frac{[\bfr \times \balpha]}{r} \,.
\ee
Here $j_1$ denotes the first-order spherical Bessel function and
$\balpha$ is the Dirac-matrix vector. In further calculations we take
into account only the first term in the power expansion of $j_1(\omega r)$,
since for the case under consideration the transition wavelength is
much larger than a typical ion size. Accordingly, the M1-transition
operator ${\bf T}^1$ can be related to the magnetic moment
operator $\bmu = e\,[\bfr \times \balpha]/2$,
\be
  {\bf T}^1 = \frac{e}{3\sqrt{2}} \, \omega \, [\bfr \times \balpha]
            = \frac{\sqrt{2}}{3} \, \omega \, \bmu \,.
  \label{t1_mu}
\ee
Utilizing the Wigner-Eckart theorem, the transition probability can be
expressed in terms of the reduced matrix element of $T^1_M$
(see, e.g., Ref.~\cite{sobelman}), which does not depend on the 
momentum projection $M$. Therefore, it is sufficient to
calculate the transition amplitude for a given projection $M$ only.
In what follows we take
$M=0$ and omit the corresponding subscript.

\begin{figure}
\includegraphics{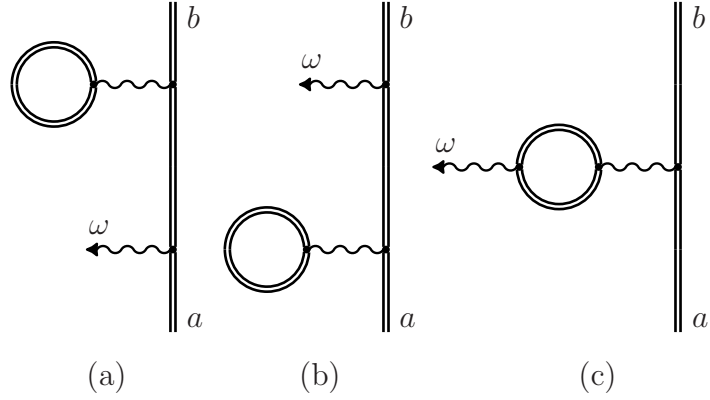}
\caption{Feynman diagrams representing the one-loop vacuum-polarization
correction to the transition amplitude.
The double line indicates the electron propagating in the external field of the
nucleus. The photon propagator is represented by the wavy line, while the
single photon emission is depicted by the wavy line with arrow.
\label{fig:vp}}
\end{figure}

\begin{figure}
\includegraphics{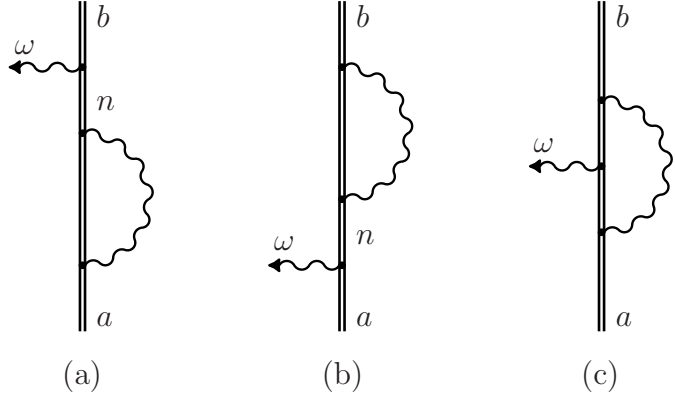}
\caption{Feynman diagrams representing the one-loop self-energy
correction to the transition amplitude.
Notations are the same as in Fig.~\ref{fig:vp}.
\label{fig:se}}
\end{figure}

In this work we focus on the one-loop QED contributions to the transition
amplitude beyond the EAMM approximation. The vacuum-polarization (VP)
and self-energy (SE) corrections, which one needs to consider, are
diagrammatically depicted in Figs.~\ref{fig:vp} and \ref{fig:se}, respectively.
The VP correction corresponding to the diagrams presented in
Figs.~\ref{fig:vp}(a) and \ref{fig:vp}(b) (the electric-loop term) has been
calculated in the Uehling potential approximation. The
diagram depicted in Fig.~\ref{fig:vp}(c) (the magnetic-loop term) has the
magnetic-interaction insertion into the VP loop. It is known, 
that the contribution of this diagram
vanishes in the Uehling approximation. The higher-orders VP
terms turn out to be rather small and can be neglected.
The remaining part of the work is devoted to the SE correction.
Here we present the formal expressions for the corresponding contributions,
which were derived at length in
Ref.~\cite{shabaev02}. 

The contributions of the diagrams depicted in Figs.~\ref{fig:se}(a) and
\ref{fig:se}(b) are conveniently divided into irreducible and reducible parts.
The reducible (``red'') contribution of the diagram depicted in
Fig.~\ref{fig:se}(a) is defined as a part in which the intermediate
state energy $\veps_n=\veps_a$, respectively $\veps_n=\veps_b$ for the diagram
presented in Fig.~\ref{fig:se}(b). The irreducible (``irr'') part is given by
the remainder.
The latter one can be written in terms of nondiagonal matrix elements of the
self-energy operator (see, for details, Ref.~\cite{shabaev02})
\be
  \label{irr}
  \Delta A^{\rm irr} = -\sqrt{\frac{2 \, \omega^3}{3\pi}}
     \left( \la b|\Sigma_{\rm R}(\veps_b)|\delta a\ra
           +\la \delta b|\Sigma_{\rm R}(\veps_a)|a\ra \right) \,,
\ee
where the perturbations to the wave functions are defined as
\be
  | \delta a \ra = \sum_n^{\veps_n\neq\veps_b}\frac{|n\ra \la n |\mu_z| a \ra}
  {\veps_b-\veps_n}, \hspace{0.5cm}
  | \delta b \ra = \sum_n^{\veps_n\neq\veps_a}\frac{|n\ra \la n |\mu_z| b \ra}
  {\veps_a-\veps_n}\,.
\ee
$\Sigma_{\rm R}(\veps)$ is the renormalized self-energy insertion,
which is related to the unrenormalized self-energy $\Sigma(\veps)$,
\be
  \label{Sigma}
  \la a |\Sigma(\veps)| b \ra = \frac{i}{2\pi}
  \int^\infty_{-\infty}\rmd E \, \sum_n
  \frac{\la a n|I(E)|n b\ra}{\veps-E-\veps_n(1-i0)}\,,
\ee
by $\Sigma_{\rm R}(\veps)=\Sigma(\veps)-\gamma^0\delta m$,
where $\delta m$ is the mass counterterm. In Eq.~(\ref{Sigma}) we use the following
notations $\alpha^\mu=(1,\balpha)$, $I(E)=e^2\alpha^\mu \alpha^\nu D_{\mu\nu}(E)$,
where $D_{\mu\nu}(E)$ is the photon propagator.
The expression for the reducible part is given by \cite{shabaev02}
\be
  \Delta A^{\rm red} = -\sqrt{\frac{\omega^3}{6\pi}} \,
  \la b|\mu_z|a\ra
  \left(\la a|\Sigma^\pr(\veps_a)|a\ra + \la b|\Sigma^\pr(\veps_b)|b\ra\right)\,,
\label{red}
\ee
where $\Sigma^\pr(\veps_a) = \rmd \Sigma(\veps) / \rmd\veps|_{\veps=\veps_a}$.
The contribution of the diagram depicted in Fig.~\ref{fig:se}(c), known as
the vertex (``ver'') term, is given by the equation \cite{shabaev02}
\be
  \Delta A^{\rm ver} = -\sqrt{\frac{2 \, \omega^3}{3\pi}}
  \frac{i}{2\pi} \int^\infty_{-\infty}\rmd E \sum_{n_{1} \, n_2}
  \frac{\la n_1|\mu_z|n_2\ra \, \la b n_2|I(E)|n_1 a\ra}
  {(\veps_b-E-\veps_{n_1}(1-i0))(\veps_a-E-\veps_{n_2}(1-i0))}\,.
\label{ver}
\ee
\begin{figure}
\includegraphics{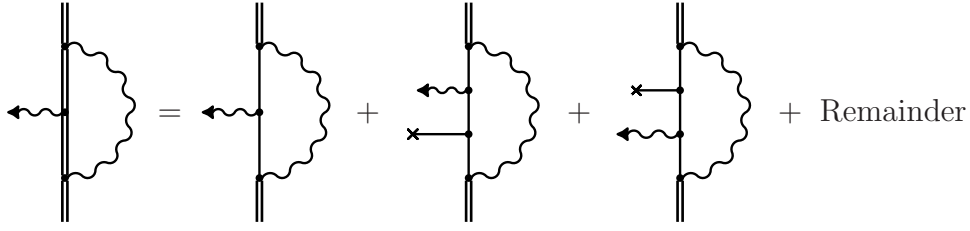}
\caption{The potential expansion of the vertex diagram.
The single line indicates the free-electron propagator and
the line ending with the cross denotes the interaction with
the field of the nucleus.
All higher-order contributions are contained in the Remainder.
\label{fig:div}}
\end{figure}

The irreducible part can be renormalized in the same manner as the ordinary
SE correction to the energy. This renormalization is well-known
and discussed in details in \cite{snyderman91,blundell91,yerokhin99}. 
The ultraviolet divergence in the vertex and reducible
contributions can be isolated by expanding the bound-electron propagator 
in terms of the interaction with the field of the nucleus.
For our purposes, it is convenient to decompose the total contribution into
zero-, one-, and many-potential terms according to the number
of interactions with the external field
\be
  \Delta A^{\rm ver}=\Delta A^{{\rm ver}(0)}
  +\Delta A^{{\rm ver}(1)}+\Delta A^{{\rm ver}(2+)}\,
\ee
and
\be
  \Delta A^{\rm red}=\Delta A^{{\rm red}(0)}
  +\Delta A^{{\rm red}(1)}+\Delta A^{{\rm red}(2+)}\,.
\ee
This expansion for the vertex diagram is schematically presented
in Fig.~\ref{fig:div}. In order to remove the divergences in the
vertex and reducible terms, we consider them together. Combining
the corresponding parts, we define
\be
  \Delta A^{{\rm vr}(i)} = \Delta A^{{\rm ver}(i)}
  +\Delta A^{{\rm red}(i)}\,,
  \qquad (i=0,1,2+) \,.
\ee
It can be shown, that the ultraviolet-divergent terms, which
are present in $\Delta A^{{\rm ver}(0)}$ and $\Delta A^{{\rm red}(0)}$,
cancel each other in $\Delta A^{{\rm vr}(0)}$.
The remaining one- and many-potential terms are ultraviolet finite.

The zero- and one-potential contributions are evaluated in
momentum space, while the many-potential term is calculated in the
coordinate space employing the partial-wave expansion.
The scheme for the separate treatment of the one-potential term 
was also used in previous g factor calculations presented in
Refs. \cite{persson97,beier00,yerokhin02,yerokhin04}.
It improves considerably the convergence of the partial-wave expansion 
in the low and middle $Z$ region.
The summarized expressions for the reducible correction are similar
to those derived for the g factor (see, Ref.~\cite{yerokhin04}).
However, for the vertex contribution there are some principal differences.
In contrast to Eq.~(4) of
Ref.~\cite{yerokhin04},
two different energies
$\veps_a$ and $\veps_b$ enter the denominator of formula (\ref{ver})
and the matrix element
$\la b n_2|I(E)|n_1 a\ra$
depends on both states $a$ and $b$. Taking these differences into account,
we derive the corresponding formulas for the $\Delta A^{{\rm ver}(0)}$
term in the Appendix.
The derivation of the formulas for the one-potential vertex
contribution is somewhat more complicated. However, taking the 
energy to be the same for both
electron propagators (e.g., $\veps_a$), the expressions
for  $\Delta A^{{\rm ver}(1)}$ can be obtained in the same
manner as for the g factor \cite{yerokhin04}.
The remaining many-potential term can be evaluated by the point-by-point
subtraction of the corresponding zero- and one-potential contributions
in the coordinate space. Consistently, we subtract the one-potential
vertex contribution with the same energy variable
in the electron propagators as it is taken in the $\Delta A^{{\rm ver}(1)}$ term
calculated
in the momentum space. Furthermore, the term  with
$\veps_{n_1}=\veps_b$ and $\veps_{n_2}=\veps_a$ in Eq.~(\ref{ver})
 has an infrared divergence, which is canceled by the corresponding term
of the reducible contribution.

For the numerical evaluation we employ the finite-basis-set method
for the Dirac equation constructed via the dual kinetic balance approach
\cite{shabaev04}. The summation of the partial-wave expansion
was performed up to $|\kappa_{\rm max}|=10$, while the remaining tail was
approximated by a least-square inverse-polynomial fitting.

%
\section{Results and discussion}
%

The one-loop QED corrections beyond the electron anomalous magnetic moment
approximation are conveniently
expressed in terms of the correction $\delta$, which is defined through
\be \label{delta}
  \Delta A_{\rm QED} = A_{\rm nr} \, (2\kappa_e+\frac{\alpha}{\pi}\delta)\,,
\ee
where $A_{\rm nr}$ is the nonrelativistic transition amplitude and
\be
  \kappa_e = \frac{\alpha}{2\pi} - 0.328\,478\,965\ldots \,
    \left ( \frac{\alpha}{\pi} \right )^2 + \cdots \,
\ee
represents the electron anomalous magnetic factor corresponding
to the EAMM term. In Table I we present our results for the one-electron
SE correction. The VP term calculated within the Uehling approximation
has been found to be negligible. The various contributions corresponding to
the SE corrections to the transition amplitude are given
in Table I. The one- and many-potential terms are represented
as the sum $\dvrom=\dvro+\dvrm$. As one can see from this table,
the occurring cancellation reduces the total value for the correction $\delta$
by an order of magnitude compared to the individual terms.
Most serious computational difficulties arise from the extrapolation of
the partial-wave expansion of the many-potential term. 
In order to estimate the error, we perform a second evaluation of $\dvrom$
without separating out the one-potential term. The difference
between the results of both calculations is taken for the uncertainty.

\begin{table}
\begin{center}
\caption{Individual contributions to the one-electron self-energy correction
expressed in terms of the various $\delta$ corrections, defined by equation (\ref{delta}).
 Numbers in parenthesis represent
error in the last digit.}
\tabcolsep10pt
\begin{tabular}{ccccc}\hline
$Z$ & $\dirr$ & $\dvrz$   & $\dvrom$     & $\delta$     \\ \hline
 16 & 0.0177  & $-$0.0124 & $-$0.0065(1) & $-$0.0012(1) \\   
 17 & 0.0195  & $-$0.0136 & $-$0.0075(1) & $-$0.0016(1) \\   
 18 & 0.0213  & $-$0.0148 & $-$0.0084(1) & $-$0.0019(1) \\   
 19 & 0.0232  & $-$0.0160 & $-$0.0094(1) & $-$0.0022(1) \\   
 20 & 0.0252  & $-$0.0172 & $-$0.0105(1) & $-$0.0025(1) \\   
 21 & 0.0272  & $-$0.0185 & $-$0.0116(1) & $-$0.0029(1) \\
 22 & 0.0292  & $-$0.0197 & $-$0.0128(1) & $-$0.0033(1) \\ \hline
\end{tabular}
\end{center}
\end{table}

The obtained results allow for an improvement of the theoretical
values of the M1-transition probabilities between the states
$(1s^2 2s^2 2p) \, ^2P_{3/2} \, - \, ^2P_{1/2}$ in B-like ions presented in \cite{tupitsyn05}.
In Table II the results of the CIDFS calculation without
the QED term $W^0$ \cite{tupitsyn05}, the improved radiative correction
$\Delta W_{\rm QED}$, the total values of the transition probability
$W_{\rm total}$, and the lifetime $\tau_{\rm total}$ are compiled.
For the transition energies we used the experimental values 
from Refs. \cite{edlen83,draganic03}. They are presented in the second column of
 Table II.
For S$^{11+}$, Cl$^{12+}$, K$^{14+}$, and Ti$^{17+}$ the uncertainties of the
values of $W_{\rm total}$ and $\tau_{\rm total}$ are determined
by the errors in the experimental transition energies.
Thus the accuracy of the total results for these ions has not been improved
in comparison with the values presented in Ref.~\cite{tupitsyn05}.
However, for Ar$^{13+}$ the experimental value for the transition energy is known
with high precision \cite{draganic03} and the uncertainty of the predictions obtained
in Ref.~\cite{tupitsyn05} is determined by the uncalculated QED terms beyond the EAMM
approximation.
The present calculation of the radiative correction $\Delta W_{\rm QED}$ has improved
the accuracy of the total values of the transition probability $W_{\rm total}$
and the lifetime $\tau_{\rm total}$ for Ar$^{13+}$ by an order of magnitude.
Furthermore, for argon ion we have added the probability of the electric-quadrupole
mode $W_{\rm E2}=0.00194$ s$^{-1}$ calculated in Coulomb gauge in Ref.~\cite{koc03}
to the total decay rate and lifetime.

In Table II, we also compare our total results with corresponding experimental data.
The disagreement with the most accurate experimental value
for Ar$^{13+}$ \cite{lapierre05,lapierre06} can be stated. The reason for this discrepancy
is still unclear for us.

\begin{table}
\begin{center}
\caption{The decay rates $W$ [s$^{-1}$] of the transition
$(1s^2 2s^2 2p) \, ^2P_{3/2} \, - \, ^2P_{1/2}$ and the lifetime $\tau$ [ms]
of the $(1s^2 2s^2 2p) \, ^2P_{3/2}$ state in B-like ions.
The transition energies are given in cm$^{-1}$.
The values of $W^0$ are taken from Ref.~\cite{tupitsyn05}. Results of the present work
$\Delta W_{\rm QED}$, $W_{\rm total}$, and $\tau_{\rm total}$ are given in columns
4 to 6. For comparison, the experimental values $\tau_{\rm expt}$ are presented in the
last column.
Numbers in parenthesis denote the estimated uncertainty.}
\tabcolsep8pt
\begin{tabular}{lcccccl}\hline
Ions      & Energy and Ref. &$W^0$ \cite{tupitsyn05} & $\Delta W_{\rm QED}$ & $W_{\rm total}$ &
                                      $\tau_{\rm total}$ & $\tau_{\rm expt}$ and Ref. \\ \hline
 S$^{11+}$&13135(1) \cite{edlen83}   &20.34481 & 0.09444 & 20.439(5)  & 48.93(1)  &                            \\
Cl$^{12+}$&17408(20) \cite{edlen83}  &47.34975 & 0.21972 & 47.57(16)  & 21.02(7)  & 21.2(6) \cite{traebert02}  \\
          &           &         &         &            &           & 21.1(5) \cite{traebert02}                 \\
Ar$^{13+}$&22656.22(1) \cite{draganic03}&104.36006& 0.48419 & 104.846(3) & 9.5378(3) & 9.12(18) \cite{moehs98} \\
          &           &         &         &            &           & 9.70(15) \cite{traebert00}                \\
          &           &         &         &            &           & 9.573(4)(5) \cite{lapierre05}             \\
 K$^{14+}$&29006(25) \cite{edlen83}&218.9394 & 1.0156  & 220.0(6)   & 4.546(12) & 4.47(10) \cite{traebert01}   \\
Ti$^{17+}$&56243(4) \cite{edlen83}&1594.714 & 7.393   & 1602.1(5)  & 0.6242(2) & 0.627(10) \cite{traebert99}   \\
\hline
\end{tabular}
\end{center}
\end{table}

%
\acknowledgments
%

Valuable conversations with O. Yu. Andreev, A. N. Artemyev, D. A. Solovyev,
and V. A. Yerokhin are gratefully acknowledged. This work was supported 
in part by RFBR (Grant No. 04-02-17574) and by INTAS-GSI (Grant No. 03-54-3604). 
 A.V.V. and G.P. acknowledge
financial support from the GSI F+E program, DFG, and BMBF.
D.A.G. acknowledges the support by the ``Dynasty'' foundation.

%
\section*{Appendix: Zero-potential vertex term}
\label{sec:zp}
%

Let us start from the momentum representation of the transition amplitude $A_{1M}$.
Referring to Eqs.~(\ref{ampl}) and (\ref{t1_mu}) it can be written as
\be
  A_{1M} = -\sqrt{\frac{\omega^3}{6\pi}} \, ie
  \int\frac{\dbp \, \dbpr}{(2\pi)^3} \, \overline{\psi}_b(\bfp) 
  \left[\balpha\times\bnabla_{\bfpr}\delta^3(\bfp-\bfpr)\right]_M
  \psi_a(\bfpr) \,,
\ee
where the gradient $\bnabla_{\bfpr}$ acts only on the $\delta$ function.
In order to obtain the zero-potential vertex term $\Delta A^{{\rm ver}(0)}$, 
we substitute $\balpha$ by the renormalized part of the free-electron vertex
operator $\bGamma_{\rm R}(\veps_b,\bfp,\veps_a,\bfpr)$
\be
  \Delta A^{{\rm ver}(0)} = -\sqrt{\frac{\omega^3}{6\pi}} \, ie
  \int\frac{\dbp \, \dbpr}{(2\pi)^3} \, \overline{\psi}_b(\bfp) 
  \left[\bGamma_{\rm R}(\veps_b,\bfp,\veps_a,\bfpr)\times
  \bnabla_{\bfpr}\delta^3(\bfp-\bfpr)\right]_z\psi_a(\bfpr) \,.
\label{verz}
\ee
In contrast to the g factor (see Eq.~(15) of Ref.~\cite{yerokhin04}),
the wave functions of the
initial ($a$) and final ($b$) states enter into Eq.~(\ref{verz}) and 
$\bGamma_{\rm R}(\veps_b,\bfp,\veps_a,\bfpr)$ has different energy arguments.
Integrating by parts and performing the integration over $\bfpr$ yields
\be
  \label{zp_vertex}
  \Delta A^{{\rm ver}(0)} &=& -\sqrt{\frac{\omega^3}{6\pi}} \, ie
  \left\{\int\frac{\dbp}{(2\pi)^3} \, \overline{\psi}_b(\bfp) \,
  \Xi(\veps_b,\veps_a,\bfp) \, \psi_a(\bfp)\right.\nonumber\\
  &&-\left.\int\frac{\dbp}{(2\pi)^3} \, \overline{\psi}_b(\bfp) 
  \left[\bGamma_{\rm R}(\veps_b,\bfp,\veps_a,\bfp)\times
  \bnabla_{\bfp}\right]_z\psi_a(\bfp)\right\} \,,
\ee
where
\be
  \Xi(\veps_b,\veps_a,\bfp) = \left[\bnabla_{\bfpr}\times
  \bGamma_{\rm R}(\veps_b,\bfp,\veps_a,\bfpr)\right]_z|_{\bfpr=\bfp} \,.
\ee
The right side of Eq.~(\ref{zp_vertex}) is naturally divided into two parts
$\Delta A^{{\rm ver}(0),1}$ and $\Delta A^{{\rm ver}(0),2}$.
Starting with the first one, it is convenient to represent the function
$\Xi(\veps_b,\veps_a,\bfp)$ in the form
\be
  \Xi(\veps_b,\veps_a,\bfp) = 4\pi i \, \alpha
  \int\frac{\dddd k}{(2\pi)^4}\frac{1}{k^2}\gamma_\sigma
  \frac{\np-\nk+m}{(p-k)^2-m^2}[\bgamma\times\bnabla_{\bfp}]_z
  \frac{\npr-\nk+m}{(p^\pr-k)^2-m^2}\gamma^\sigma \,
\ee
with $p=(\veps_b,\bfp)$, $p^\pr=(\veps_a,\bfp)$, and $\np=p_\mu\gamma^\mu$,
respectively.
Using the commutation identity for the $\gamma$ matrices, we get
\be
  \Xi(\veps_b,\veps_a,\bfp) &=& \frac{\alpha}{4i\pi^3}
  \int\frac{\dddd k}{k^2}\frac{1}{\left[(p-k)^2-m^2\right]
  \left[(p^\pr-k)^2-m^2\right]}\Bigl\{
  \gamma_\sigma(\np-\nk+m)[\bgamma\times\bgamma]_z\gamma^\sigma\nonumber\\
  &&+2 \, \gamma_\sigma\frac{(\np-\nk+m)(\npr-\nk-m)}{(p^\pr-k)^2-m^2}
  [\bgamma\times(\bfp-\bfk)]_z\gamma^\sigma\Bigr\} \,.
\ee
Expressing the integration over the loop momenta $k$ in terms
of the integrals over the Feynman parameters, one can derive
the formula
\be
  \Xi(\veps_b,\veps_a,\bfp) &=& \frac{\alpha}{\pi}
  \Bigl\{i\gamma_0\gamma_5\gamma_z(C_0+C_{11}+C_{12})\np
  -\Bigl[\npr(A_0-A_1)\np-(A_0+3A_1)\nonumber\\
  &&+2p^2(A_{11}-A_{21})
  +2p^{\pr \, 2}(A_{12}-A_{22})-4p^\mu p^\pr_\mu A_{23}\Bigr]
  [\bgamma\times\bfp]_z\Bigr\} \,,
\ee
which coincides with the corresponding equation in calculations of
the g factor, if one considers $\veps_a=\veps_b$. Here the Feynman integrals are
determined as
\be
  C_0 = \int_0^1 \frac{\rmd y}{(yp+(1-y)p^\pr)^2}(-\ln \, X)\,,
\ee
\be
  \Biggl(\begin{array}{c}
   C_{11}\\
   C_{12}\end{array}\Biggr) = \int_0^1 \frac{\rmd y}{(yp+(1-y)p^\pr)^2}
   \Biggl(\begin{array}{c}
   y\\
   1-y\end{array}\Biggr)(1-Y\ln \, X)\,,
\ee
\be
  A_0 = \int_0^1 \rmd x\rmd y \, \frac{(1-x)(1-y)}{Z^2}\,,
\ee
\be
  A_1 = \int_0^1 \rmd x\rmd y \, \frac{x(1-x)(1-y)}{Z^2}\,,
\ee
\be
  \Biggl(\begin{array}{c}
   A_{11}\\
   A_{12}\end{array}\Biggr) = \int_0^1 \rmd x\rmd y \,
                              \frac{x(1-x)(1-y)}{Z^2}
   \Biggl(\begin{array}{c}
   y\\
   1-y\end{array}\Biggr)\,,
\ee
\be
  \left(\begin{array}{c}
   A_{21}\\
   A_{22}\\
   A_{23}\end{array}\right) = \int_0^1 \rmd x\rmd y \,
                              \frac{x^2(1-x)(1-y)}{Z^2}
   \left(\begin{array}{c}
   y^2\\
   (1-y)^2\\
   y(1-y)\end{array}\right)\,,
\ee
and
\be
  X = 1 + \frac{1}{Y}\,,
\ee
\be
  Y = \frac{1-yp^2-(1-y)p^{\pr \, 2}}{(yp+(1-y)p^\pr)^2}\,,
\ee
$Z=x[yp+(1-y)p^\pr]^2+1-yp^2-(1-y)p^{\pr \, 2}$.

To carry out the angular integration for the transition
under consideration $2p_{3/2} - 2p_{1/2}$
we employ the following results for basic integrals ($\mu=1/2$)
\be
  \label{ang1}
  \int\rmd\Omega_\bfp \chid_{\kappa_1\mu}(\hp)\sigma_z\chi_{\kappa_2\mu}(\hp)=
  \Biggl\{\begin{array}{c}
   -2\sqrt{2}/3\\
   0\end{array}
   \begin{array}{c}
   {\rm for}\\
   {\rm for}\end{array}
   \begin{array}{c}
   \kappa_1=1,\\
   \kappa_1=-1,\end{array}
   \begin{array}{c}
   \kappa_2=-2\,,\\
   \kappa_2=2\,,\end{array}
\ee
\be
  \label{ang2}
  \int\rmd\Omega_\bfp \chid_{\kappa_1\mu}(\hp)[\bsigma\times\hp]_z\chi_{\kappa_2\mu}(\hp)=
  \Biggl\{\begin{array}{c}
   -\sqrt{2}/3 \, i\\
    \sqrt{2}/3 \, i\end{array}
   \begin{array}{c}
   {\rm for}\\
   {\rm for}\end{array}
   \begin{array}{c}
   \kappa_1=-1,\\
   \kappa_1=1,\end{array}
   \begin{array}{c}
   \kappa_2=-2\,,\\
   \kappa_2=2\,,\end{array}
\ee
where $\hat{\bfp}=\bfp/|\bfp|$, $\chi_{\kappa\mu}(\hp)$ is the spherical spinor,
and $\bsigma$ denotes the vector of Pauli matrices.
Finally, for the first part $\Delta A^{{\rm ver}(0),1}$ we obtain
\be
  \label{zp_vertex_1}
  \Delta A^{{\rm ver}(0),1} &=& -\sqrt{\frac{\omega^3}{3\pi}}
  \frac{\alpha e}{24\pi^4}
  \int_0^\infty\rmd\rp \, \rp^2\Biggl\{
  -2(C_0+C_{11}+C_{12})(\veps_b g_b g_a + \rp g_b f_a)\nonumber\\
  &&+\rp\Bigl[(\veps_a\veps_b-\rp^2)(A_0-A_1-4A_{23})
   -(A_0+3A_1)+2(\veps_b^2-\rp^2)\nonumber\\
   &&\times(A_{11}-A_{21})
   +2(\veps_a^2-\rp^2)(A_{12}-A_{22})\Bigr]
   (g_b f_a - f_b g_a)\nonumber\\
  &&-\rp^2(\veps_a-\veps_b)(A_0-A_1)(g_b g_a - f_b f_a)
  \Biggr\}\,,
\ee
where $\rp=|\bfp|$, $g_a(\rp)$ and $f_a(\rp)$ are the upper and lower radial
components of the wave function in the momentum representation, respectively.

The second term $\Delta A^{{\rm ver}(0),2}$ can be calculated similarly.
Using the expression for the free-electron vertex function \cite{yerokhin99}
and employing in addition to Eq.~(\ref{ang2}) the following
angular integrals
\be
  \label{ang3}
  \int\rmd\Omega_\bfp \chid_{\kappa_1\mu}(\hp)[\hp\times\bnabla_{\Omega_\bfp}]_z\chi_{\kappa_2\mu}(\hp)=
  \Biggl\{\begin{array}{c}
   \sqrt{2}/3 \, i\\
   0\end{array}
   \begin{array}{c}
   {\rm for}\\
   {\rm for}\end{array}
   \begin{array}{c}
   \kappa_1=1,\\
   \kappa_1=-1,\end{array}
   \begin{array}{c}
   \kappa_2=-2\,,\\
   \kappa_2=2\,,\end{array}
\ee
\be
  \label{ang4}
  \int\rmd\Omega_\bfp \chid_{\kappa_1\mu}(\hp)[\bsigma\times\bnabla_{\Omega_\bfp}]_z\chi_{\kappa_2\mu}(\hp)=
  \Biggl\{\begin{array}{c}
   -2\sqrt{2}/3 \, i\\
   \sqrt{2} \, i\end{array}
   \begin{array}{c}
   {\rm for}\\
   {\rm for}\end{array}
   \begin{array}{c}
   \kappa_1=-1,\\
   \kappa_1=1,\end{array}
   \begin{array}{c}
   \kappa_2=-2\,,\\
   \kappa_2=2\,,\end{array}
\ee
where $\bnabla_{\Omega_\bfp}$ is the angular part of the gradient,
we have
\be
  \label{zp_vertex_2}
  \Delta A^{{\rm ver}(0),2} &=& -\sqrt{\frac{\omega^3}{3\pi}}
  \frac{\alpha e}{96\pi^4}
  \int_0^\infty\rmd\rp \, \rp^2\Biggl\{
  (A-\veps_a\veps_b D+\rp^2 D)(g_b f_a^\pr - f_b g_a^\pr\nonumber\\ 
  &&+\frac{3}{\rp}g_b f_a-\frac{2}{\rp}f_b g_a)
  -\rp D(\veps_a-\veps_b)(f_b f_a^\pr - g_b g_a^\pr + \frac{3}{\rp} f_b f_a - \frac{2}{\rp} g_b g_a )
  \nonumber\\
  &&+(\veps_b B+2\veps_b D+\veps_a C+4D) g_b g_a
  +\rp(B+2D+C) f_b g_a \Biggr\}\,,
\ee
where $g_a^\pr(\rp)=\rmd g_a(\rp)/\rmd \rp$, $f_a^\pr(\rp)=\rmd f_a(\rp)/\rmd \rp$.
Here the set of coefficients are expressed in terms
of the Feynman integrals as
\be  
  A &=& C_{24}-2+(\veps_b^2-\rp^2)C_{11}+(\veps_a^2-\rp^2)C_{12}
       +4(\veps_a\veps_b-\rp^2)(C_0+C_{11}+C_{12})\nonumber\\
       &&-2C_0+C_{11}+C_{12}\,,
\ee
\be  
  B = -4(C_0+2C_{11}+C_{12}+C_{21}+C_{23})\,,
\ee
\be  
  C = -4(C_0+C_{11}+2C_{12}+C_{22}+C_{23})\,,
\ee
\be  
  D = 2(C_0+C_{11}+C_{12})\,,
\ee
and
\be  
  C_{24} = -\int_0^1\rmd y \, \ln(y^2(\veps_a-\veps_b)^2
           -y(\veps_a-\veps_b)^2+1)\,.
\ee
The total result for the zero-potential vertex contribution
is the sum of the corresponding terms from Eqs.~(\ref{zp_vertex_1})
and (\ref{zp_vertex_2}).
%

%

\begin{thebibliography}{99}
\clearpage
%
\bibitem{moehs98}
D. P. Moehs and D. A. Church, Phys. Rev. A {\bf 58}, 1111 (1998).
%
\bibitem{traebert99}
E. Tr\"abert, G. Gwinner, A. Wolf, X. Tordoir,
and A. G. Calamai, Phys. Lett. A {\bf 264}, 311 (1999).
%
\bibitem{traebert00}
E. Tr\"abert, P. Beiersdorfer, S. B. Utter,
G. V. Brown, H. Chen, C. L. Harris, P. A. Neill,
D. W. Savin, and A. J. Smith, Astrophys. J. {\bf 541}, 506 (2000).
%
\bibitem{traebert01}
E. Tr\"abert, P. Beiersdorfer, G. V. Brown,
H. Chen, E. H. Pinnington, and D. B. Thorn,
Phys. Rev. A {\bf 64}, 034501 (2001).
%
\bibitem{traebert02}
E. Tr\"abert, P. Beiersdorfer, G. Gwinner,
E. H. Pinnington, and A. Wolf, Phys. Rev. A {\bf 66}, 052507 (2002).
%
\bibitem{lapierre05}
A. Lapierre, U. D. Jentschura, J. R. Crespo L\'opez-Urrutia, J. Braun,
G. Brenner, H. Bruhns, D. Fischer, A. J. Gonz\'alez Mart\'inez, Z. Harman,
W. R. Johnson, C. H. Keitel, V. Mironov, C. J. Osborne, G. Sikler,
R. Soria Orts, V. M. Shabaev, H. Tawara, I. I. Tupitsyn, J. Ullrich,
and A. V. Volotka,
Phys. Rev. Lett. {\bf 95}, 183001 (2005).
%
\bibitem{lapierre06}
A. Lapierre, J. R. Crespo L\'opez-Urrutia, J. Braun,
G. Brenner, H. Bruhns, D. Fischer, A. J. Gonz\'alez Mart\'inez,
V. Mironov, C. J. Osborne, G. Sikler, R. Soria Orts, H. Tawara, J. Ullrich,
V. M. Shabaev, I. I. Tupitsyn, and A. V. Volotka,
to be published.
%
\bibitem{tupitsyn05}
I. I. Tupitsyn, A. V. Volotka, D. A. Glazov, V. M. Shabaev, G. Plunien,
J. R. Crespo L\'opez-Urrutia, A. Lapierre, and J. Ullrich,
Phys. Rev. A. {\bf 72}, 062503 (2005).
%
\bibitem{indelicato96}
P. Indelicato, Phys. Rev. Lett. {\bf 77}, 3323 (1996).
%
\bibitem{derevianko98}
A. Derevianko, I. M. Savukov, W. R. Johnson, and
D. R. Plante, Phys. Rev. A {\bf 58}, 4453 (1998).
%
\bibitem{indelicato04}
P. Indelicato, V. M. Shabaev, and A. V. Volotka,
Phys. Rev. A {\bf 69}, 062506 (2004).
%
\bibitem{sapirstein04}
J. Sapirstein, K. Pachucki, and K. T. Cheng,
Phys. Rev. A {\bf 69}, 022113 (2004).
%
\bibitem{shabaev05a}
V. M. Shabaev, K. Pachucki, I. I. Tupitsyn, and V. A. Yerokhin,
Phys. Rev. Lett. {\bf 94}, 213002 (2005).
%
\bibitem{shabaev05b}
V. M. Shabaev, I. I. Tupitsyn, K. Pachucki, G. Plunien, and V. A. Yerokhin,
Phys. Rev. A. {\bf 72}, 062105 (2005).
%
\bibitem{shabaev98}
V. M. Shabaev,
Can. J. Phys. {\bf 76}, 907 (1998).
%
\bibitem{blundell97}
S. A. Blundell, K. T. Cheng, and J. Sapirstein,
Phys. Rev. A {\bf 55}, 1857 (1997).
%
\bibitem{persson97}
H. Persson, S. Salomonson, P. Sunnergren, and I. Lindgren,
Phys. Rev. A {\bf 56}, R2499 (1997).
%
\bibitem{beier00}
T. Beier, I. Lindgren, H. Persson, S. Salomonson, P. Sunnergren, H. H\"affner,
and N. Hermanspahn,
Phys. Rev. A {\bf 62}, 032510 (2000).
%
\bibitem{yerokhin02}
V. A. Yerokhin, P. Indelicato, and V. M. Shabaev,
Phys. Rev. Lett. {\bf 89}, 143001 (2002).
%
\bibitem{yerokhin04}
V. A. Yerokhin, P. Indelicato, and V. M. Shabaev,
Phys. Rev. A {\bf 69}, 052503 (2004).
%
\bibitem{sobelman}
I. I. Sobelman, {\it Atomic Spectra and Radiative Transitions},
(Springer, New York, 1979).
%
\bibitem{shabaev02}
V. M. Shabaev, Phys. Rep. {\bf 356}, 119 (2002).
%
\bibitem{snyderman91}
N. J. Snyderman,
Ann. Phys. (N.Y.) {\bf 211}, 43 (1991).
%
\bibitem{blundell91}
S. A. Blundell and N. J. Snyderman,
Phys. Rev. A {\bf 44}, R1427 (1991).
%
\bibitem{yerokhin99}
V. A. Yerokhin and V. M. Shabaev,
Phys. Rev. A {\bf 60}, 800 (1999).
%
\bibitem{shabaev04}
V. M. Shabaev, I. I. Tupitsyn, V. A. Yerokhin, G. Plunien, and G. Soff,
Phys. Rev. Lett. {\bf 93}, 130405 (2004).
%
\bibitem{edlen83}
B. Edl\'en, Phys. Scripta {\bf 28}, 483 (1983).
%
\bibitem{draganic03}
I. Dragani\'c, J. R. Crespo L\'opez-Urrutia, R. DuBois,
S. Fritzsche, V. M. Shabaev, R. Soria Orts, I. I. Tupitsyn,
Y. Zou, and J. Ullrich, Phys. Rev. Lett. {\bf 91}, 183001 (2003).
%
\bibitem{koc03}
K. Koc, J. Phys. B {\bf 36}, L93 (2003).
%
\end{thebibliography}
\end{document}